\documentclass[12pt]{article}
\usepackage{graphicx}
\usepackage{cite}
\usepackage{color}
\newcommand{\bm}[1]{\mbox{\boldmath $#1$}}
\begin{document}
\title{Scalar and axial-vector mesons}
\author{
Eef van Beveren\\
{\normalsize\it Centro de F\'{\i}sica Te\'{o}rica}\\
{\normalsize\it Departamento de F\'{\i}sica, Universidade de Coimbra}\\
{\normalsize\it P-3004-516 Coimbra, Portugal}\\
{\small http://cft.fis.uc.pt/eef}\\ [.3cm]
\and
George Rupp\\
{\normalsize\it Centro de F\'{\i}sica das Interac\c{c}\~{o}es Fundamentais,
Instituto Superior T\'{e}cnico}\\
{\normalsize\it Universidade T\'{e}cnica de Lisboa, Edif\'{\i}cio Ci\^{e}ncia,
P-1049-001 Lisboa, Portugal}\\
{\small george@ajax.ist.utl.pt}\\ [.3cm]
{\small PACS number(s): 14.40.-n, 14.40.Cs, 14.40.Ev, 14.40.Lb}
}

\maketitle

\begin{abstract}
Almost thirty years ago, Penny G. Estabrooks asked
``Where and what are the scalar mesons?'' \cite{PRD19p2678}.
The first part of her question can now be confidently
responded \cite{ZPC30p615}. However, with respect to the ``What''
many puzzles remain unanswered.
Scalar and axial-vector mesons form part of a large family
of mesons. Consequently, though it is useful to pay them
some extra attention, there is no point in discussing them
as isolated phenomena. The particularity of structures in
the scattering of --- basically --- pions and kaons with zero
angular momentum is the absence of the centrifugal
barrier, which allows us to `see' strong interactions at
short distances. Experimentally observed differences and
similarities between scalar and axial-vector mesons on the
one hand, and other mesons on the other hand, are very
instructive for further studies.
Nowadays, there exists an abundance of theoretical approaches
towards the mesonic spectrum, ranging from confinement models
of all kinds, i.e., glueballs, and quark-antiquark, multiquark and
hybrid configurations, to models in which only mesonic degrees
of freedom are taken into account. Nature seems to come out somewhere
in the middle, neither preferring pure bound states, nor
effective meson-meson physics with only coupling constants
and possibly form factors. As a matter of fact, apart from a
few exceptions, like pions and kaons, Nature does not allow us
to study mesonic bound states of any kind, which is equivalent
to saying that such states do not really exist. Hence, instead
of extrapolating from pions and kaons to the remainder of the
meson family, it is more democratic to consider pions and kaons
mesonic resonances that happen to come out below the lowest
threshold for strong decay. Nevertheless, confinement is an
important ingredient for understanding the many regularities
observed in mesonic spectra. Therefore, excluding quark degrees
of freedom is also not the most obvious way of describing mesons
in general, and scalars and axial-vectors in particular.
\end{abstract}
\clearpage

\section{Introduction}

Since all known mesonic resonances have quantum numbers (spin $J$,
parity $P$, charge-conjugation parity $C$, flavour/isospin)
which agree with the quantum numbers of a confined pair of a quark
and an antiquark, it is reasonable to set out a description in terms
of a confined two-particle system, like the harmonic oscillator (HO)
\cite{ccbarHOjpegs}.
However, the experimental mass spectrum of mesonic resonances
does not agree with the spectrum of
the ordinary HO \cite{HOjpegs}:
masses from the HO and from experiment differ substantially,
while, furthermore, many HO states are not seen in experiment.

Recent discoveries
\cite{PRL89p102001},
\cite{HEPEX0607082}
seem to indicate that the present experimental spectrum for mesonic
resonances is still very incomplete,
but masses are usually well determined.
Hence, the HO classification of states may work well,
but it fails to reproduce the masses accurately.
However, reproducing the data
is not only reproducing resonance positions and possibly widths.
One must also reproduce the full scattering data of experiment,
even at energies where no resonances show up,
and in all possible scattering channels.
Consequently, it is not at all clear that the HO does not work well
for confinement, as long as the effects of all possible meson loops
have not yet been accounted for \cite{PRD21p772}.
Moreover, upon exploring the Weyl-conformal-invariance property of QED/QCD,
one obtains Anti-DeSitter (AdS) confinement
\cite{NPS244p82}
with flavour-independent HO-like level spacings of spectra.
The linear-plus-Coulomb type of confinement is in the AdS approach
obtained from one-gluon exchange at short distances \cite{LNP211p331}.
Furthermore, in lattice-based results, one also seems to observe
a HO-like spectrum \cite{HEPPH0603046}, but the common practice of fitting
lattice parameters to the ground states of the spectrum leads to too large
level splittings at higher energies
\cite{HEPLAT0510060}.

Unquenching the confinement spectrum has been studied by various groups,
and for a variety of different confinement mechanisms
\cite{AP123p1,Cargese75p305,PRD21p772,AIPCP717p665,PRD50p6855}.
The procedure usually amounts to the inclusion of meson loops in a $q\bar{q}$
description, or, equivalently,
the inclusion of quark loops in a model for meson-meson scattering,
resulting in resonance widths,
central masses that do not coincide with the pure confinement spectrum,
mass shifts of bound states,
resonance line-shapes that are very different from the usual Breit-Wigner
ones, threshold effects and cusps.
In particular, it should be mentioned that
mass shifts are large and negative for the ground states
of the various flavour configurations \cite{PRD27p1527}.
Unquenching the lattice is still in its infancy, at least for the light
scalars, as we conclude from Ref.~\cite{HEPLAT0510066}.
However, its effects should not be underestimated.
Hence, ground-state levels of quenched approximations
for $q\bar{q}$ configurations in relative $S$-waves
must be expected to come out far above the experimental masses.

From pure 2-body harmonic-oscillator confinement we expect to find
18 $c\bar{c}$ states in the mass region of 4.0$\pm$0.2 GeV,
according to the quantum numbers given in Table~\ref{ccbar4GeV}.
Nevertheless, we encounter only three states
with established quantum numbers in Ref.~\cite{PLB592p1}, i.e.,
$\psi (3770)$ at 3.77 GeV,
$\psi (4040)$ at 4.04 GeV and
$\psi (4160)$ at 4.16 GeV.
\begin{table}[htbp]
\caption[]{\small $c\bar{c}$ states expected in the mass region
4.0$\pm$0.2 GeV.}
\label{ccbar4GeV}
\begin{center}
\begin{tabular}{||c||c|c|c|c|c||}
\hline\hline & & & & & \\ [-7pt]
$J^{PC}$ & $0^{-+}$ & $1^{--}$ & $1^{+-}$ & $0^{++}$ & $1^{++}$\\
& & & & & \\ [-7pt]
\hline & & & & & \\ [-7pt]
degeneracy & 1 & 2 & 1 & 1 & 1\\
& & & & & \\ [-7pt]
\hline\hline & & & & & \\ [-7pt]
$J^{PC}$ & $2^{++}$ & $2^{-+}$ & $2^{--}$ & $3^{--}$ & $3^{+-}$\\
& & & & & \\ [-7pt]
\hline & & & & & \\ [-7pt]
degeneracy & 2 & 1 & 1 & 2 & 1\\ [3pt]
& & & & & \\ [-7pt]
\hline\hline & & & & & \\ [-7pt]
$J^{PC}$ & $3^{++}$ & $4^{++}$ & $4^{-+}$ & $4^{--}$ & $5^{--}$\\
& & & & & \\ [-7pt]
\hline & & & & & \\ [-7pt]
degeneracy & 1 & 1 & 1 & 1 & 1 \\ [3pt]
\hline\hline
\end{tabular}
\end{center}
\end{table}
The discovery of possibly four more $c\bar{c}$ states in this mass region
\cite{PRL91p262001}
starts filling the many gaps in the experimental $c\bar{c}$ spectrum.
But for unquenching HO confinement, one also needs some detailed
experimental information on charmed decay modes.
However, the only experiment which addresses this issue
\cite{PRL37p255}
dates from {\bf 1977}, and reports results that are at odds
with naive expectations \cite{PRL37p398}.

In this context it is opportune to quote the remark of E.~Swanson
in his excellent review on the newly discovered states \cite{HEPPH0601110}:
``{\it It is worth noting that attempts to unquench the quark model
are fraught with technical difficulty
and a great deal of effort is required
before we can be confident in the results of any model.}''

\section{Coupled channels.}

Using coupled-channel techniques, one can simulate unquenching.
In Refs.~\cite{PRD17p3090,PRD21p772}, quark-pair creation was modelled
by coupling $q\bar{q}$ confinement and meson-meson scattering channels.
Yu.~Kalashnikova \cite{PRD72p034010} applied this method to describe
the new $X(3872)$ state \cite{PRL91p262001} by coupling
$c\bar{c}$ with $J^{P}=1^{+}$ to $D$-meson pairs ($D\bar{D}$,
$D\bar{D}^{\ast}$, $D^{\ast}\bar{D}^{\ast}$, $D_{s}\bar{D}_{s}$,
$D_{s}\bar{D}_{s}^{\ast}$ and $D_{s}^{\ast}\bar{D}_{s}^{\ast}$).
In Refs.~\cite{PRD17p3090,PRD21p772}, the same technique was applied
to bound states below the lowest threshold for strong decay.
Hence, in this philosophy,
such states contain components of virtual meson pairs.
C.~Albertus described in his talk a method to actually observe the virtual
$B^{\ast}\pi$ component of the $B$ meson in semileptonic $B\to\pi\ell\nu$
decay \cite{CAlbertus}.

In other approaches, the composition of confinement channels
is not {\it a priori} \/known,
since their effects are replaced by resonance-pole exchanges
\cite{ZPC68p647,PRD54p1991,NUCLTH0307039}.
These methods have the advantage that the difficult issue of confinement
does not have to be addressed when analysing scattering data.
In particular in the scalar-meson sector it has shown to be a convincing
strategy.
\vspace{10pt}

\noindent
{\bf \bm{D_{s0}^{\ast}(2317)} and its first radial excitation.}
Confinement dictates the quantum numbers of the states,
and indicates the mass region where one may expect to observe them.
Fine structure follows from additional interactions,
which may even generate `dynamically' extra resonances.

When in the model of Ref.~\cite{PRD27p1527} one determines the mass
of the lowest $c\bar{s}$ state in a relative $P$-wave, one obtains 2.545 GeV
for pure HO confinement, so even larger than the mass of 2.48 GeV predicted
by S.~Godfrey and N.~Isgur \cite{PRD32p189}.
But under the creation of a non-strange quark pair this system couples
to $D(c\bar{n})+K(n\bar{s})$, which has its threshold at 2.363 GeV.
Unquenching the $c\bar{s}$ state, by allowing it to couple to $DK$
\cite{PRL91p012003},
brings the ground-state mass of the full system down
to 2.32 GeV, exactly where it has been found in experiment
\cite{PRL90p242001,STosi}.
Similar results have been obtained by D.~S.~Hwang and D.-W.~Kim
\cite{PLB601p137}, and by Yu.~Simonov and J.~A.~Tjon \cite{PRD70p114013}.

There exist many alternative explanations for the mass of the
$D_{s0}^{\ast}(2317)$.
Exploiting the full Dirac structure of confined $q\bar{q}$ states,
T.~Matsuki and collaborators got 2.446 GeV in Ref.~\cite{PRD56p5646}
and, after refining their parameters, 2.330 GeV in Ref.~\cite{HEPPH0605019}.
Applying the resonating-group method to a chiral-symmetric quark model,
P.~Bicudo obtained short-range meson-meson attraction and so
$DK$ molecules \cite{NPA748p537}.
This interesting result partly confirmed the description in
Ref.~\cite{PRL91p012003}.
In the latter work, the $D_{s0}^{\ast}(2317)$ was considered
a two-component object: 1) a pure $c\bar{s}$ state in a relative $P$-wave;
2) an $S$-wave $DK$ state. Hence, it certainly contains a virtual $DK$
molecular-like component.

Th.~Mehen and R.~Springer \cite{PRD72p034006} concluded that
including counterterms is critical for fitting current data
of scalar and axial-vector charmed mesons with
one-loop chiral corrections.
By imposing simultaneously the constraints from chiral symmetry
and heavy-quark spin symmetry on effective theories of heavy-light hadrons,
M.~Nowak and J.~Wasiluk \cite{APPB35p3021} advocated that
$D_{s0}^{\ast}(2317)$ and $D_{s1}(2460)$ must be viewed as
the chiral doublers of $D_{s}$ and $D_{s}^{\ast}$, which was contested by
P.~Bicudo \cite{PRD74p036008}.
A.~Zhang \cite{PRD72p017902} found that
the slopes of Regge trajectories decrease
with increasing quark mass, and predicted 2.35 GeV
for the mass of the missing $J^{P}=1^{+}$ $D$ meson.
We certainly endorse his recommendation that
{\it ``predicted states should be searched for and more}
\/(strong) {\it decay modes should be detected.''}

Since the mass of the $D_{s0}^{\ast}(2317)$ ends up below the threshold
\begin{table*}[htbp]
\caption[]{\small The experimentally observed light positive-parity mesons.}
\label{pospar}
\begin{center}
\begin{tabular}{||c||cc|c|cccc||}
\hline\hline & & & & & & & \\ [-7pt]
& \multicolumn{2}{c|}{$I=1$} & $I=\frac{1}{2}$ &
\multicolumn{4}{c||}{$I=0$} \\
& & & & & & & \\ [-7pt]
\hline & & & & & & & \\ [-7pt]
$0^{+}$ &
$a_{0}(1450)$ & &
$K_{0}(1430)$ & $f_{0}(1370)$ & $f_{0}(1500)$ & & \\
& & &
$K_{0}(1980)$ & $f_{0}(1710)$ & $f_{0}(2020)$ & & \\
& & & & $f_{0}(2200)$ & & & \\
& & & & & & & \\ [-7pt]
\hline & & & & & & & \\ [-7pt]
$1^{+}$ &
$a_{1}(1260)$ & $b_{1}(1235)$ &
$K_{1}(1270)$ &
$f_{1}(1285)$ & $f_{1}(1420)$ & $h_{1}(1170)$ & $h_{1}(1380)$ \\
&
$a_{1}(1640)$ & &
$K_{1}(1400)$  &
$f_{1}(1510)$ & & $h_{1}(1595)$ & \\
& & &
$K_{1}(1650)$ & & & & \\
& & & & & & & \\ [-7pt]
\hline & & & & & & & \\ [-7pt]
$2^{+}$ &
$a_{2}(1320)$ & &
$K_{2}(1340)$ & $f_{2}(1270)$ & & $f_{2}(1430)$ & \\
&
$a_{2}(1700)$ & & &
$f_{2}(1525)$ & $f_{2}(1565)$ & $f_{2}(1640)$ & $f_{2}(1810)$ \\
& & &
$K_{2}(1980)$ &
$f_{2}(1910)$ & $f_{2}(1950)$ & $f_{2}(2010)$ & $f_{2}(2150)$ \\
& & &
& $f_{2}(2300)$ & $f_{2}(2340)$ & & \\ [3pt]
\hline\hline
\end{tabular}
\end{center}
\end{table*}
of the lowest OZI-allowed \cite{OZI} decay mode (i.e.\ $DK$),
it represents a bound state in this specific selection of decay channels,
which we consider the most important. Consequently, the
$D_{s0}^{\ast}(2317)$ may be represented by a pole on the real energy axis.
The pole representing the first radial excitation of the $c\bar{s}$
system in a relative $P$-wave comes out well above the $DK$ threshold in
the model of Ref.~\cite{IJTPGTNO11p179}.
Consequently, this pole has a negative imaginary part
\cite{HEPPH0606110,HEPPH0610188}, which gives rise to the width
of the radial excitation of the $D_{s0}^{\ast}(2317)$.
In Ref.~\cite{HEPPH0606110}, two poles were found,
one at 2.32 GeV and a second at $(2.85-i0.024)$ GeV,
representing the ground state and the first radial
excitation of the $J^{P}=0^{+}$ $c\bar{s}$ system, respectively.
Experiment \cite{HEPEX0607082} reported a $c\bar{s}$ structure at 2.86 GeV,
with precisely the same line-shape as the theoretical prediction
\cite{HEPPH0610188},
and being compatible with $J^{P}=0^{+}$ quantum numbers.
However, an alternative explanation exists \cite{HEPPH0607245}.

But in Ref.~\cite{HEPPH0606110} an additional pole showed up
in the scattering amplitude of the model of Ref.~\cite{IJTPGTNO11p179}.
Its theoretical position was reported at $(2.78-i0.23)$ GeV.
\begin{figure}[htbp]
\vspace{-20pt}
\begin{center}
\begin{tabular}{l}
\resizebox{0.4\textwidth}{!}{\includegraphics{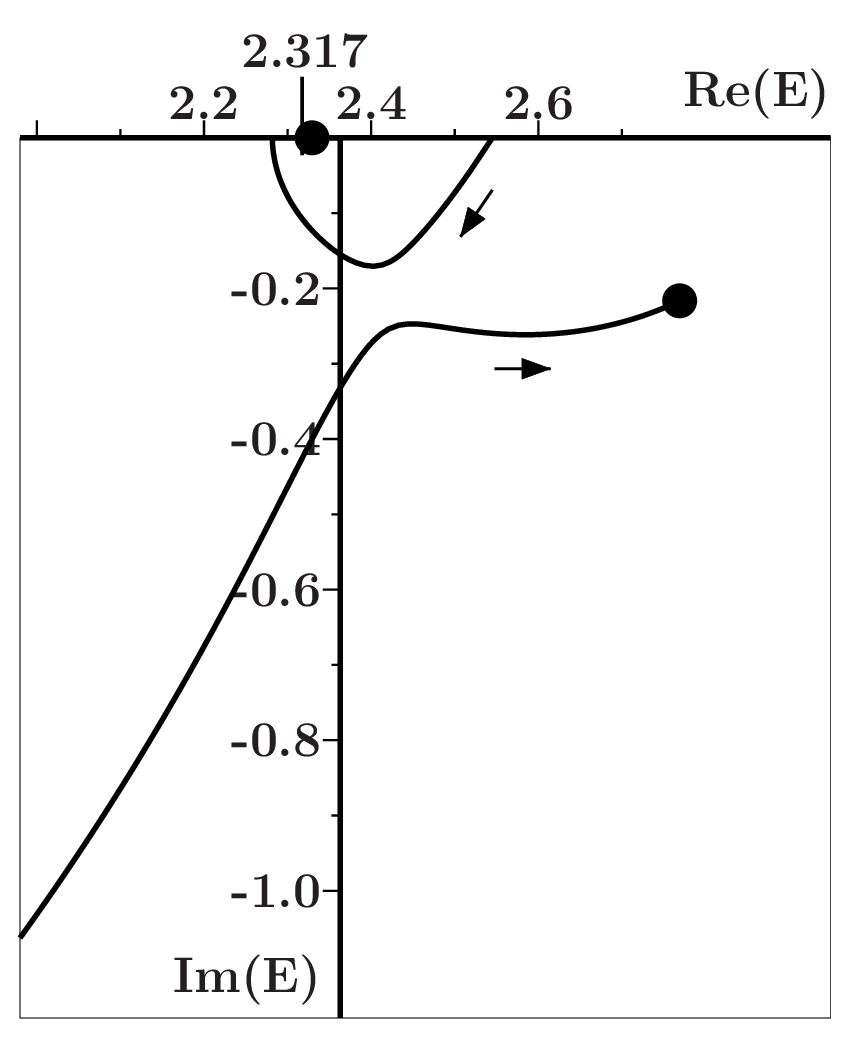}}
\end{tabular}
\end{center}
\caption[]{Trajectories of poles in the $DK$ $S$-wave
scattering amplitude for the model
of Ref.~\cite{HEPPH0606110}, as a function of the amount of unquenching.
In the quenched approximation, the dynamically generated pole has negative
infinite imaginary part, whereas the confinement ground state comes out at
$\sqrt{s}=2.454$ GeV on the real axis.
The arrows indicate how the poles move when unquenching increases.
The model's physical values are indicated by dots.
The imaginary axis is drawn at the $DK$ threshold.}
\label{Ds0poles}
\end{figure}
The appearance of such {\it dynamically generated} \/poles is no surprise,
since similar (at that time `unexpected') poles had been observed
and reported \cite{ZPC30p615} for the model of Ref.~\cite{PRD27p1527},
explaining well the existence of the $J^{P}=0^{+}$ phenomena below 1.0 GeV,
as the natural consequence of combining confinement and quark-pair creation.

In Fig.~\ref{Ds0poles} we show the trajectories of
the two lowest-lying poles in the scattering matrix
for increasing $c\bar{s}$-$DK$ coupling.
The BABAR collaboration reported in Ref.~\cite{HEPEX0607082}
on the possible existence of a broad $c\bar{s}$ resonance, which might correspond
to the dynamically generated pole.

E.~Kolomeitsev and M.~Lutz \cite{PLB582p39} obtained a bound state at
2303 MeV for a coupled $DK$-$D_ {s}\eta$ system from their non-linear chiral
$SU(3)$ Lagrangian.

Four-quark configurations have been proposed to explain the mass of the
$D_{s0}^{\ast}(2317)$, namely by L.~Maiani {\em et al.}
\/\cite{PoSHEP2005p105}, by H.~Cheng and W.~Hou \cite{PLB566p193},
and by K.~Terasaki \cite{HEPPH0309119}.
The latter author argued in his talk \cite{KTerasaki}
why the doubly-charged partners have not been seen by BABAR \cite{STosi}.
In Ref.~\cite{PLB599p55}, it was concluded that the $q\bar{q}$ picture
is not adequate for charmed scalar mesons, based on the conflict between theory
and experiment for chiral-loop corrections to the mass differences between the
scalar and pseudoscalar heavy-light mesons:
``{\it the unitarised  meson model of Ref.~\cite{ZPC30p615}
or the 4-quark picture for the scalar mesons~\cite{PLB566p193,HEPPH0309119}
may be a useful remedy in explaining the scalar states
containing one heavy quark.''}
\clearpage

\section{Positive-parity mesons.}

In Table~\ref{pospar}, we show the experimental spectrum of light
positive-parity mesons. Only the $f_2$ states allow for comparison with HO
confinement. We therefore assume that the states in the first two $f_2$ columns
contain mostly non-strange $q\bar{q}$ pairs, and in the next two
predominantly $s\bar{s}$ pairs.
Furthermore, the quark pair may come
in a relative $P$-wave
(1\raisebox{1.0ex}{\scriptsize st} and
3\raisebox{1.0ex}{\scriptsize rd} column), or
in an $F$-wave
(2\raisebox{1.0ex}{\scriptsize nd} and
4\raisebox{1.0ex}{\scriptsize th} column).
From the values for the central mass positions as given in
Ref.~\cite{PLB592p1}, we collect in Table~\ref{massdiff}
mass differences for a selected set of $f_{2}$ states.
\begin{table}[htbp]
\caption[]{\small The experimentally \cite{PLB592p1}
observed mass differences for a selected set of
light positive-parity mesons.}
\label{massdiff}
\begin{center}
\begin{tabular}{||c|c||}
\hline\hline & \\ [-7pt]
states & mass difference \\
& \\ [-7pt]
\hline & \\ [-7pt]
$m\left( f_{2}(1910)\right) - m\left( f_{2}(1525)\right)$ &
0.39 $\pm$ 0.01 GeV \\ [5pt]
$m\left( f_{2}(2300)\right) - m\left( f_{2}(1910)\right)$ &
0.38 $\pm$ 0.03 GeV \\ [5pt]
$m\left( f_{2}(1950)\right) - m\left( f_{2}(1565)\right)$ &
0.40 $\pm$ 0.02 GeV \\ [5pt]
$m\left( f_{2}(2340)\right) - m\left( f_{2}(1950)\right)$ &
0.39 $\pm$ 0.04 GeV \\ [5pt]
$m\left( f_{2}(2010)\right) - m\left( f_{2}(1640)\right)$ &
0.38 $\pm$ 0.05 GeV \\ [5pt]
$m\left( f_{2}(2150)\right) - m\left( f_{2}(1810)\right)$ &
0.34 $\pm$ 0.02 GeV \\ [5pt]
\hline\hline
\end{tabular}
\end{center}
\end{table}
For HO confinement one has a level spacing of 0.38 GeV \cite{PRD27p1527},
which agrees well with the splittings in Table~\ref{massdiff}.
The degeneracy lifting of $P$-wave and $F$-wave
is some 40 $\pm$ 10 MeV for non-strange
(1\raisebox{1.0ex}{\scriptsize st} and
2\raisebox{1.0ex}{\scriptsize nd} $f_2$ column in Table~\ref{pospar}),
and about 165 $\pm$ 20 MeV for strange
(3\raisebox{1.0ex}{\scriptsize rd} and
4\raisebox{1.0ex}{\scriptsize th} column).
But the non-degenerate ground states $f_{2}(1270)$ and $f_{2}(1430)$
do seem too high in mass with respect to the splittings in
Table~\ref{massdiff}.
In order to understand that, we have to return to Fig.~\ref{Ds0poles}.

The two trajectories shown in Fig.~\ref{Ds0poles} come close
to each other for certain values of the $c\bar{s}$-$DK$ coupling.
Upon a variation of one other model parameter, this becomes a saddle point.
Depending on the value of this parameter, the trajectories may interchange.
In that case the end points are connected differently,
making the $D_{s0}^{\ast}(2317)$ the dynamically generated state,
whereas the other pole then seems to stem from the confinement ground state.
This is actually what appears to happen for the light positive-parity
ground-state mesons and makes them move up in energy when unquenching is
turned on.
For the scalar mesons, those states correspond to the
$f_{0}(1370)$, $f_{0}(1500)$, $K_{0}^{\ast}(1430)$ and $a_{0}(1450)$.
The dynamically generated poles correspond
to the lower-lying scalar mesons \cite{ZPC30p615}.

\section{The light scalar mesons.}

\begin{figure}[htbp]
\begin{center}
\begin{tabular}{ll}
\resizebox{0.4\textwidth}{!}{\includegraphics{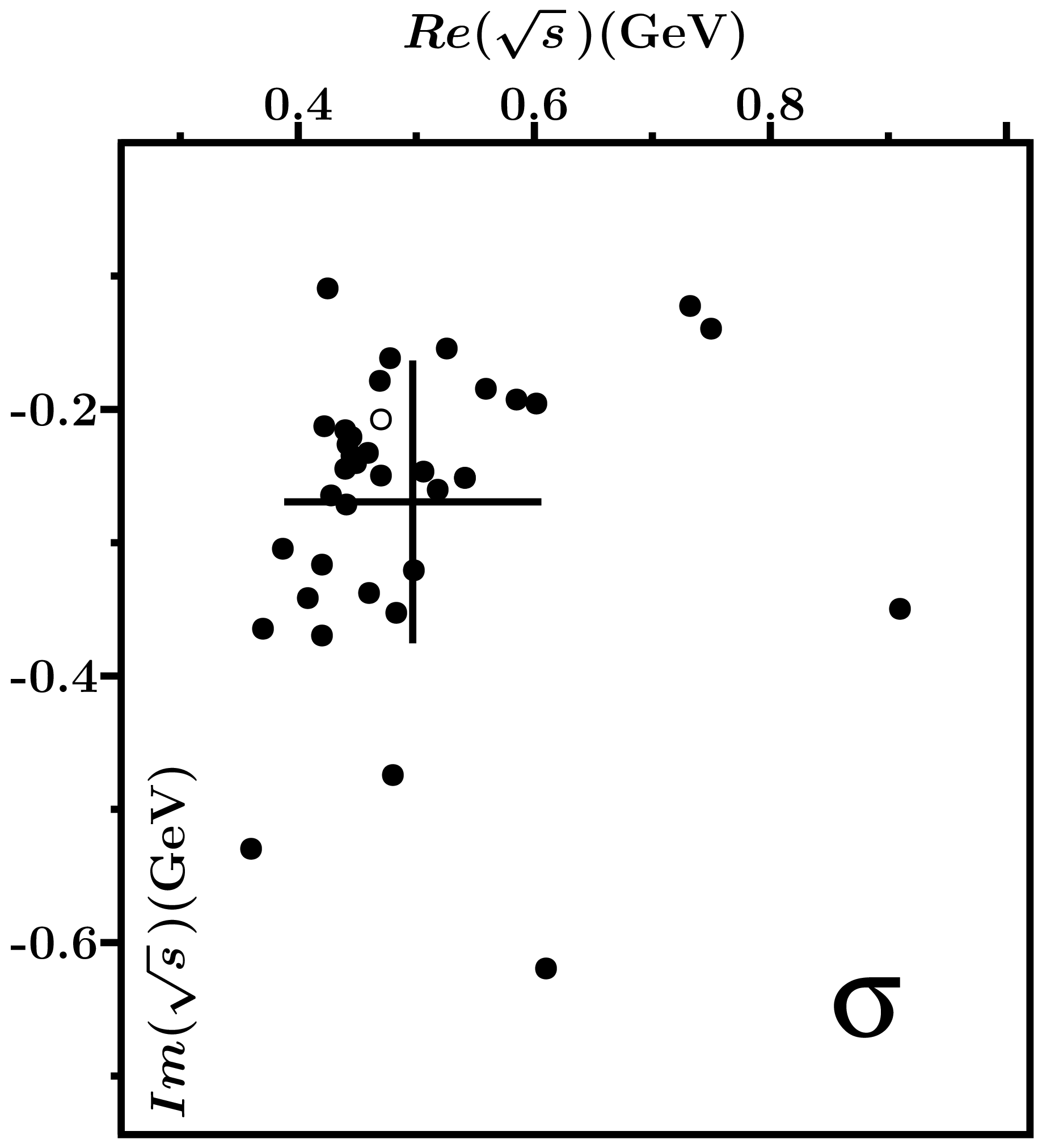}} &
\resizebox{0.4\textwidth}{!}{\includegraphics{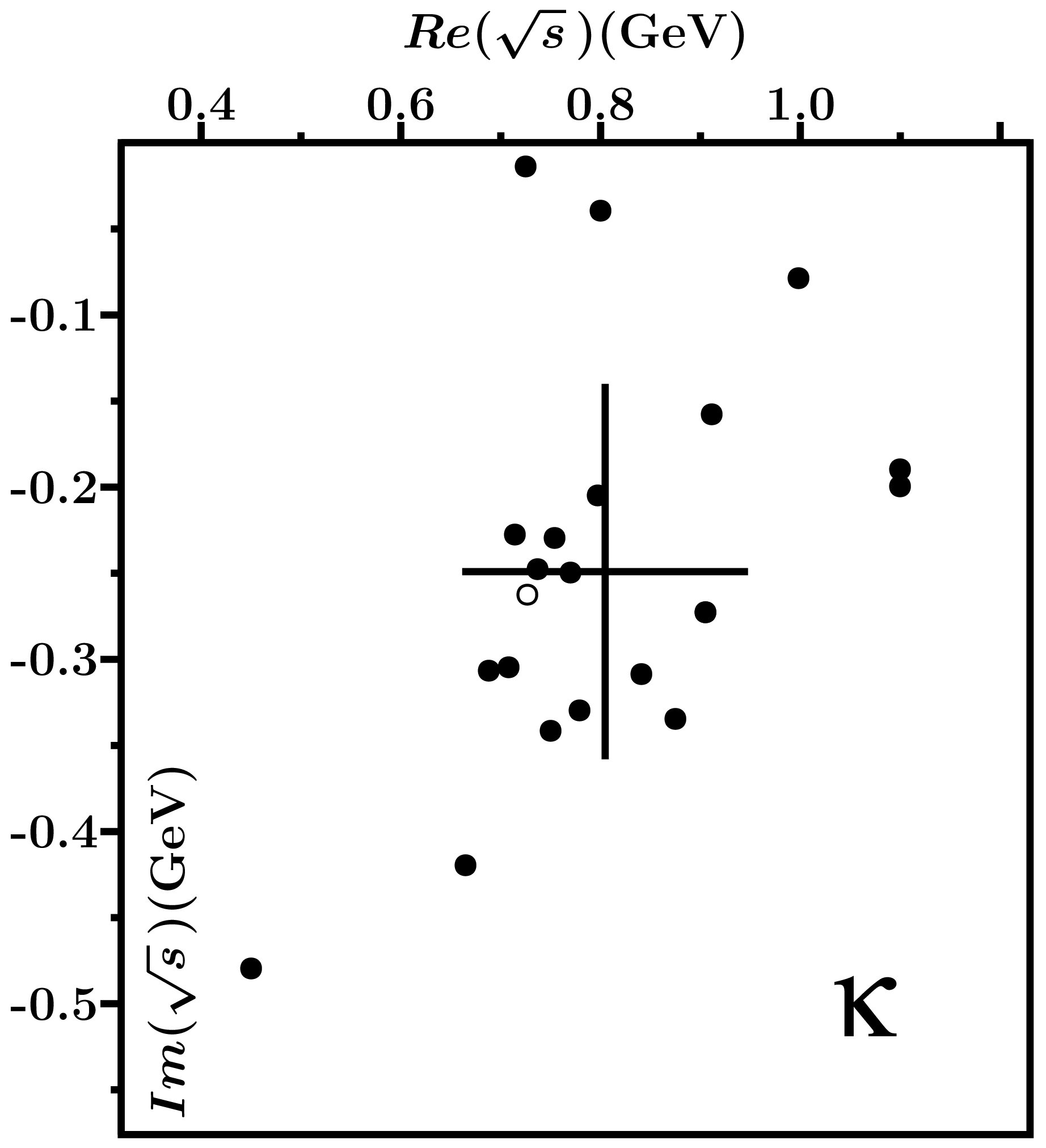}}
\end{tabular}
\end{center}
\caption[]{Pole positions found in the literature,
for the two most disputed scalar mesons: $f_{0}(600)$ and $K_{0}^{\ast}(800)$.
From the above collection of real and imaginary parts,
we find on average for the $f_{0}(600)$ pole position
(497 $\pm$ 109)$-i$(269 $\pm$ 106) MeV,
and for the $K_{0}^{\ast}(800)$
(805 $\pm$ 143)$-i$(249 $\pm$ 109) MeV.
The figures are taken from {\tt
http:\-//cft.\-fis.\-uc.\-pt/\-eef/\-sigkap.htm}
where, moreover, all references are given.
The open circles correspond to the scattering poles in Ref.~\cite{ZPC30p615}.}
\label{sigkap}
\end{figure}

The scalar mesons have been a source of inspiration and controversy
since the 1960s.
Particularly the $\epsilon$(600) and $\kappa$(900) \cite{RMP45pS1},
nowadays called $f_ {0}$(600) (or simply $\sigma$) and $K_{0}^{\ast}$(800)
\cite{PLB592p1}, respectively, disappeared from the `Particle Listings' in
the late 1970s \cite{RMP52pS1}.
Their nature is still not settled, nor for their nonet partners
$f_ {0}$(980) and $a_ {0}$(980).
But it seems that we are converging towards the idea
that low-energy scattering data are best described by
a nonet of scalar $S$-matrix poles \cite{JPG28pR249}.
In Fig.~\ref{sigkap} we collect some of the pole positions
encountered in the literature for the $\sigma$ and the $\kappa$.

As early as in 1968, P.~Roy \cite{PR168p1708},
using current-algebra sum rules, estimated that a $\kappa$ meson might exist,
with a central mass at 725 MeV and a width of $\ge\!28\pm2$ MeV.
Later, J.~Basdevant and B.~Lee \cite{PRD2p1680}
determined, for various values of $f_{\pi}$,
the $\sigma$ mass and width within the $\sigma$ model.
For $f_{\pi}=95$ MeV, they found
$M_{\sigma}=425$ MeV and $\Gamma_{\sigma}=220$ MeV.
In Ref.~\cite{NPB60p233}, a Lagrangian for vector-meson exchange was
employed to determine masses 460 MeV and 665 MeV,
and widths 675 MeV and 840 MeV, for
the $\sigma$ and the $\kappa$, respectively.
In 1982, M.~D.~Scadron \cite{PRD26p239} extended to the entire scalar nonet
the dynamical spontaneous breakdown of chiral symmetry for the QCD
quark theory, elaborated with R.~Delbourgo in Ref.~\cite{PRL48p379},
resulting in $\sigma$(750), $\kappa$(800), $f_{0}$(980)
and $a_{0}$(985), with respective widths of
280 MeV, 80 MeV, 24 MeV and 58 MeV.

In the meantime, based on the phenomenology of te\-tra\-quarks
($qq\bar{q}\bar{q}$),
R.~L.~Jaffe \cite{PRD15p267} also had come to the conclusion
that ``{\it there should be a very broad kaon-pion enhancement
at roughly 900 MeV}.''
The reason for considering multi-quark configurations
was based on the observation that naive confinement models,
taylormade for $c\bar{c}$ and $b\bar{b}$,
produce spectra for positive-parity mesons
comparable to the spectrum shown in Table~\ref{pospar},
with no sign of the light scalar mesons, and, furthermore, in order to explain
the degeneracy of $f_{0}$(980) and $a_{0}$(980).
In Ref.~\cite{ZPC30p615} (1986), however, it was shown
that the coupled-channel approach, representing quark-pair creation,
is capable of describing the spectra of heavy and light quarks with one
set of parameters, and that, as a bonus,
the full light scalar nonet pops up without even being anticipated,
with $f_{0}$(980) and $a_{0}$(980) almost degenerate, and having widths
comparable to experiment
(the $\sigma$ and $\kappa$ results for Ref.~\cite{ZPC30p615}
are shown in Fig.~\ref{sigkap}).
The effects of $S$-wave thresholds have been studied in
Refs.~\cite{LNP211p331,HEPPH0207022,HEPPH0608102}.

Nevertheless, many years later, the J\"{u}lich group did not find
a $\kappa$ pole in their meson-exchange model \cite{PRD52p2690},
nor did N.~T\"{o}rnqvist and M.~Roos \cite{PRL76p1575},
nor the $K$-matrix analysis of A.~Anisovich and A.~Sarantsev
\cite{PLB413p137}.
Moreover, in order to explain part of the scalar-meson nonet with
the lowest-lying $0^{++}$ glueball,
P.~Minkowski and W.~Ochs argued that rather a distorted nonet
containing the $f_{0}$(1500) as a partner \cite{EPJC9p283}
is consistent with what can be expected theoretically.
The issue culminated in a paper by S.~Cherry and M.~Pennington
\cite{NPA688p823}, which claimed in the title
``{\it There is no $\kappa$(900)}'',
followed by Ref.~\cite{PRD65p114010},
in which M.~Boglione and M.~Pennington even argued that the $\kappa$ pole
might be found on the real axis, below the $K\pi$ threshold.

Many more works have appeared on the matter of the light scalar mesons
and, in particular, on the existence and position of the $\kappa$ pole.
The Ishidas, with K.~Takamatsu and T.~Tsuru,
applying the method of interfering Breit-Wigner amplitudes
to a reanalysis of the $K\pi$ $S$-wave phase shifts,
found evidence for the existence of a $\kappa$(900)
\cite{PTP98p621}.
D.~Black, A.~Fariborz, F.~Sannino and J.~Schechter,
studying meson-meson scattering with the use of
an effective chiral Lagrangian,
found a $\kappa (900)$, together with $\sigma (560)$, $a_0(980)$ and
$f_0(980)$ \cite{PRD59p074026}.
But they also remarked that fitting the light scalars
into a nonet pattern suggests that the underlying structure is
closer to diquark-antidiquark than to quark-antiquark.
M.~Volkov and V.~Yudichev proposed that the $f_0$(1500) should
be composed mostly of the scalar glueball \cite{EPJC10p223}.
Y.~Dai and Y.~Wu \cite{EPJC39pS1} concluded, using a non-linear effective
chiral Lagrangian for meson fields, obtained from integrating out the quark
fields by using a finite regularisation method, that
the lightest nonet of scalar mesons, which appear as the chiral partners of
the nonet of pseudoscalar mesons, should be composite Higgs bosons with masses
below the chiral-symmetry-breaking scale $\Lambda_{\chi} \sim 1.2$ GeV.
H.~Q.~Zheng and collaborators \cite{NPA733p235} showed that the
$\kappa$ resonance exists, if the scattering-length parameter in the
$I\!=\!1/2$ and $J\!=\!0$ channel does not deviate much from its value
predicted by chiral perturbation theory.
T.~Kunihiro {\em et al.} \cite{NPPS129p242} reported on very heavy $\kappa$
mesons in unquenched lattice calculations.
Y.~Oh and H.~Kim \cite{PRC74p015208} proposed that the scalar $\kappa(800)$
meson may play an important role in $K^*$ photoproduction, and in particular
that the parity asymmetry can separate the
$\kappa$-meson contribution in $K^*$ photoproduction.

A breakthrough came from the $E791$ experiment,
with a clear $\kappa(800)$ signal \cite{PRL89p121801}.
But the analysis of production data is far from trivial, and
seems to be still under study \cite{AReis}.
However, in Ref.~\cite{PRD71p054030} J.~Oller showed that scattering data
for $\pi\pi$ and $K\pi$ \cite{NPB133p490}
are in perfect agreement with the more recent production data
\cite{PRL86p770,PRL89p121801},
using the unitarised chiral perturbation method,
earlier developed with E.~Oset and J.~Pel\'{a}ez
\cite{PRD59p074001}.
Recently, also the BES collaboration found evidence for
$\kappa$-meson production, i.e., in
$J/\psi\to\bar{K}^{\ast\,0}(892)K^{+}\pi^{-}$ \cite{PLB633p681}.
In his analysis of the {\it combined} \/LASS, E791 and BES data, D.~V.~Bugg
concluded that $K\pi$ is fitted well with a $\kappa$ pole at
$(750\pm30)-i(342\pm60)$ MeV and the usual $K_{0}$(1430) resonance
\cite{PLB632p471}.

In Ref.~\cite{PRD71p116002}, a scalar nonet $\kappa (1045)$,
$\sigma (600)$, $a_0(873)$ and $f_0(980)$ was obtained,
using the extended three-flavour NJL model, which has no quark confinement,
and including 4- and 6-quark interactions.
In Ref.~\cite{PLB634p48}, it was shown that one needs to include
at least also 8-quark interactions in the NJL model in order to obtain
globally stable vacuum solutions.
This has, moreover, a considerable effect
on the mass of the $\sigma$ resonance \cite{HEPPH0607066}.
Finally, Ref.~\cite{PRD74p037501} demonstrated how,
in the model of Ref.~\cite{PRL91p012003},
by just varying the flavour-mass parameters
(and at the same time the related threshold masses),
the poles of the universal $S$-matrix
transform into one another, thus
relating e.g.\ $\kappa$(800) to $D_{s0}^{\ast}(2317)$
via a continuous process.

\section{Conclusions.}

In the past decade, great progress has been made in the theoretical
understanding of the mesonic sector of hadronic resonances, in particular
of the light-scalar nonet. The efforts in setting up and carefully analysing
new experiments have greatly contributed to this achievement.
Nevertheless, it should be recognised that still a lot of additional data are
needed and must be unravelled, in particular on resonance positions and
strong-decay branching ratios, before we can hope for a more complete
understanding of strong interactions.

\section*{Acknowledgments}

We thank the organisers of the conference for their hospitality
and their efforts to bring together many specialists in a great variety
of areas. It has been a very inspiring and fruitful week.
This work was partly supported by the
{\it Funda\c{c}\~{a}o para a Ci\^{e}ncia e a Tecnologia}
of the {\it Minist\'{e}rio da
Ci\^{e}ncia e do Ensino Superior} \/of Portugal,
under contract
POCI/\-FP/\-63437/\-2005.

\newcommand{\pubprt}[4]{#1 {\bf #2}, #3 (#4)}
\def\AIPCP{AIP Conf.\ Proc.}
\def\AP{Annals Phys.}
\def\APPB{Acta Phys.\ Polon.\ B}
\def\EPJC{Eur.\ Phys.\ J.\ C}
\def\IJTPGTNO{Int.\ J.\ Theor.\ Phys.\ Group Theor.\ Nonlin.\ Opt.}
\def\JPG{J.\ Phys.\ G}
\def\LNP{Lect.\ Notes Phys.}
\def\MPLA{Mod.\ Phys.\ Lett.\ A}
\def\NPA{Nucl.\ Phys.\ A}
\def\NPB{Nucl.\ Phys.\ B}
\def\NPPS{Nucl.\ Phys.\ Proc.\ Suppl.}
\def\NPS{Nature Phys.\ Sci.}
\def\PLB{Phys.\ Lett.\ B}
\def\PR{Phys.\ Rev.}
\def\PRC{Phys.\ Rev.\ C}
\def\PRD{Phys.\ Rev.\ D}
\def\PRL{Phys.\ Rev.\ Lett.}
\def\PREP{Phys.\ Rept.}
\def\PTP{Prog.\ Theor.\ Phys.}
\def\RMP{Rev.\ Mod.\ Phys.}
\def\ZPC{Z.\ Phys.\ C}

\end{document}